\documentclass[a4paper,11pt,aps,nofootinbib]{revtex4}
\usepackage{amsmath}
\usepackage{bbm}
\usepackage{graphicx,amssymb,bm,latexsym,color,epsf}
\usepackage{breqn}
\usepackage[colorlinks]{hyperref} 

\newcommand{\be}{\begin{equation}}
\newcommand{\ee}{\end{equation}}
\newcommand{\bea}{\begin{eqnarray}}
\newcommand{\eea}{\end{eqnarray}}
\newcommand{\G}{\Gamma}
\newcommand{\tr}{\mathrm{tr}}
\newcommand{\Tr}{\mathrm{Tr}}

\newcommand{\La}{\Lambda}
\newcommand{\reg}{\rho_{k, \Lambda}(s)}
\newcommand{\ZL}{Z_\Lambda(\phi)}
\newcommand{\VL}{V_\Lambda(\phi)}
\newcommand{\DS}{\delta S^{(2)}}

\newcommand{\NAG}{\texttt{NAG} }

\makeatletter
\let\cat@comma@active\@empty
\makeatother
\begin{document}

\title{On Exact Proper Time Wilsonian RG Flows}
\author{Alfio Bonanno}
\email{alfio.bonanno@inaf.it}
\affiliation{INAF, Osservatorio Astrofisico di Catania, via S.Sofia 78, I-9 5123 Catania, Italy}
\affiliation{ INFN, Sezione di Catania, via S. Sofia 64, I-95123,Catania, Italy}

\author{Stefan Lippoldt}
\email{s.lippoldt@thphys.uni-heidelberg.de}
\affiliation{Institut f\"ur Theoretische Physik, Universit\"at Heidelberg, Philosophenweg 16, 69120 Heidelberg, Germany}

\author{Roberto Percacci}
\email{percacci@sissa.it}
\affiliation{ SISSA, via Bonomea 265, I-34136 Trieste}
\affiliation{ INFN, Sezione di Trieste, Italy}

\author{Gian Paolo Vacca}
\email{vacca@bo.infn.it}
\affiliation{INFN, Sezione di Bologna, via Irnerio 46, I-40126 Bologna, Italy}

\begin{abstract}
We discuss the possibility to define exact RG equations for a UV regulated Wilsonian action based on a proper time (PT) regulator function. 
We start from a functional mapping which shows how each particular flow equation (and RG scheme) is associated to infinitely many scale dependent field redefinitions, which are related to specific coarse-graining procedures. On specializing to a sub-family of one parameter PT regulators we briefly analyze few results for the Ising Universality class in three dimensions, obtained within a second order truncation in the derivative expansion of the Wilsonian action.

\end{abstract} 
\maketitle



\section{Introduction}


When a generating functional in a quantum field theory is defined
by means of a functional integral with a built-in cutoff,
its derivative with respect to the cutoff gives rise to a Renormalization Group (RG) flow.
The original idea is due to Wilson \cite{Wilson_1975},
who defined an action $S_\Lambda$ depending on an UV cutoff $\Lambda$
in such a way that when $S_\Lambda$ is used in a functional integral
with a cutoff $\Lambda$,
it gives the same partition function independently on the choice of 
$\La$. In other words, the physical predictions at low energy are unchanged: 
\begin{equation}
 Z=\int [d\varphi]_\La e^{-S_\La[\varphi]} \quad , \quad \La \frac{\rm d}{{\rm d} \La} Z =0\,.
\end{equation}
The subscript $\La$ in the measure means that momentum integrals
are to be cut off in the UV at $\La$.
There is no unique way to define the cutoff, but all
these Wilsonian RG flows can be shown to satisfy some kind of
differential equation, usually called an Exact RG Equation (ERGE).
One of the oldest forms of the ERGE is due to Wegner and Houghton
\cite{Wegner:1972ih}.
Better-known in the context of particle physics is the
Polchinski equation \cite{Polchinski_1984}, giving
the flow of the interaction part of the action.

A different class of equations arise when we consider the
generating functional of 1-particle irreducible correlators,
the Effective Action (EA) $\Gamma$.
In this case one introduces in the functional integral an IR cutoff,
usually called $k$, and the functional $\Gamma_k$ is called
the Effective Average Action (EAA) \cite{average}.
It satisfies an exact RG equation known as the Wetterich equation
\cite{wetterich1, Morris:1993qb, morris1,morris2}.
Employing various approximation schemes, these exact equations 
can be used to derive many results in condensed matter,
statistical and particle physics.
For reviews, see \cite{morrisrev,Bagnuls,Berges,Delamotte,Rosten}.
For applications to gauge theories see
\cite{reuterwetterich, Gies, Pawlowski}.

At one loop, in the Schwinger proper time (PT) representation of
the effective action, one can introduce the cutoff 
(either UV or IR or both) in the integral 
over the proper time, rather than the integral over momenta
\cite{Floreanini:1995aj,Liao:1994fp}.
The resulting ``proper time RG equation'' for the EAA
is known not to be exact~\cite{Litim:2001hk,Litim:2001ky,Litim:2002xm}, but nevertheless
has been applied to various problems
from statistical physics \cite{Zappala:2002nx,Bonanno:2004pq}
to gravity \cite{Bonanno:2004sy}, and gives comparable results
to the various forms of the ERGE.
This is generally attributed to the fact that the one loop
approximation becomes exact when one considers integration
over infinitesimal momentum shells \cite{Wegner:1972ih}.
Recently, a PT-like equation for the Wilsonian action,
rather than the EAA,
has been introduced in \cite{deAlwis:2017ysy}:
\begin{equation}
 \La \frac{d S_\La[\varphi]}{d \La} = \Tr \ e^{-S^{(2)}_\La[\varphi] /\La^2}\,.
\label{eq:PTexp}
\end{equation}
It has been applied to issues of quantum gravity.
Unlike previous PTRG equations, this one is claimed to be exact.
We stress that this would not be in contrast with the previous statement for the 1PI functional (EAA) because now the equation holds for a Wilsonian action $S_\Lambda$,
which has to be used in the functional integral of the low momentum modes.
In this paper we will examine by different means 
a more general family of equations which can characterize the RG flow of the Wilsonian action $S_\La$ and, 
as a simple quantitative test, 
apply it to derive the scaling exponents of the Ising universality class.

\section{Wilsonian PTRG equations}\label{sec2}

Consider a general family of PT flows
\begin{equation} 
\label{eq:genPT_flow}
\La \frac{d S_\La[\varphi]}{d \La}
= \frac{1}{2} \Tr \int_{0}^{\infty} \! \frac{ds}{s} \, 
r_\La(s)
 e^{-s S^{(2)}_{\La}[\varphi]} \ .
\end{equation}
Here $s$ is the Schwinger proper time,
\be
\label{eq:cut}
r_\La(s)=\La \frac{d \rho_{k,\La}}{d \La}
\ee
and $\reg$ is a suitably normalized regulator function depending on an IR scale $k$ 
and an UV scale $\Lambda$.
It goes to zero for $s<1/\La^2$ (UV regularization)
and $s>1/k^2$ (IR regularization), and is roughly equal to one
in between.
Choosing the regulator function
\begin{equation} \label{thetareg}
\reg=\theta(s-1/\Lambda^{2}) - \theta(s-1/k^{2})\ ,
\end{equation}
one obtains de Alwis' equation (\ref{eq:PTexp}).
%
We will now discuss other choices of regulator.
Let us consider the following 1-parameter family of regulators ($m\geq 0$),
\begin{equation}
\rho_{k, \La}(s ; m) = \frac{\G(m , m s k^{2}) - \G(m , m s \La^{2})}{\G(m)} .
\label{eq:spectral_rho_gamma}
\end{equation}
Using the properties of the incomplete Gamma function we obtain
\begin{equation} \label{eq:spectral_der_rho_gamma}
r_{\La}(s ; m)
= \frac{2}{\G(m)}(m s \La^{2})^{m} e^{- m s \La^{2}}\, .
\end{equation}
Note that this object is actually dependent on the 
UV cutoff $\La$ only, justifying the notation (\ref{eq:cut}).
We shall call this the ``A-scheme''.
Inserting this choice of regulator in (\ref{eq:genPT_flow}) and performing the Mellin transform we get the following RG
equation for the Wilsonian action $S_\La$,
\begin{equation} \label{eq:new-ERGE}
 \La \frac{d}{d \La} S_{\La}[\varphi] = \Tr \left( \frac{m \La^{2}}{S^{(2)}_{\La}[\varphi] + m \La^{2}} \right)^{m} .
\end{equation}
Note that for $m = 1$ this looks like the Wetterich Equation with a massive regulator (which can define a flow for the IR regulated effective average action in low dimensionality).
In the LPA approximation, the specific value $m = d/2 + 1$ 
gives essentially the same flow for the
potential as the Wetterich equation with an optimized regulator.
One should not be misled by these analogies, since $S_\La$
has to be interpreted only as a Wilsonian action 
(to appear inside a functional integral) and not 
as an IR-regulated generator of the 1PI correlators
\cite{Litim:2002hj}.

Note also that for $m \rightarrow \infty$ the scale derivative of $\reg$ becomes a Dirac delta distribution, or equivalently,
one has in~(\ref{eq:new-ERGE}) the representation of an exponential, so the flow equation reduces again to the one given
in Eq.~(\ref{eq:PTexp}).

The cutoffs \eqref{thetareg} only depend on $s$, $k$ and $\La$.
It is also possible to introduce in the PT regulator
some field dependence. 
This is related to the idea of ``spectral adjustment''.
Let us start by considering the derivative expansion of a 
generic Wilsonian action $S_{\La}(\varphi)$, with 
$S^{(2)}_{\La}(\varphi) = Z_{\La}(\varphi)(-\Box) + \cdots$.
Here $\varphi$ has to be slowly varying on length scales
of order $p^{-1}$, where $p$ are the momenta appearing in $-\Box$.
In equation \eqref{eq:genPT_flow}, the regulator $\rho(s)$
suppresses the part of the integral with $s<1/\La^2$, 
while the exponential suppresses modes with $Z_{\La}p^{2}>1/s$.
Altogether the suppression occurs for $Z_{\La}p^{2}>\La^{2}$.
We see that if $Z_\La$ is sizable, the cutoff is not directly related to the spacetime scale of the fluctuation: this scale
is modulated by a flowing field-dependent dynamical factor $Z_\La$.
Let us consider instead a more ``spectrally adjusted'' 
scheme obtained using a proper time regulator $\rho(Z_{\La} s)$.
In this case the regulator suppresses the part of the integral 
with $s < \frac{1}{\La^{2} Z_{\La}}$, 
which corresponds to a suppression of the modes with
$p^{2} > \La^{2}$.
This is closer to the original intent of
cutting off the modes at the sliding scale $\La$.
One observes that it is in the spectrally adjusted scheme that the two steps in the renormalization procedure, 
namely (1) coarse-graining and (2) rescaling, are tuned to each other as desired for a comparison along the flow.

If we ``spectrally adjust'' in this fashion the cutoff
\eqref{eq:spectral_rho_gamma}
\begin{equation}
\rho_{k, \La}(s;m) = \frac{\G(m , m s\, Z_k(\phi)k^{2}) 
- \G(m , m s\, Z_\La(\phi)\La^{2})}{\G(m)} .
\label{eq:spectral_rho_gammaB}
\end{equation}
we obtain
\begin{equation} \label{eq:spectral_der_rho_gammaB}
r_{\La}(s ; m)
= \frac{2}{\G(m)}\left( 1 + 
\frac{1}{2} \La \frac{d}{d \La} \log Z_\La(\phi)\right)
(m s Z_\La(\phi)\La^{2})^{m} e^{- m s Z_\La(\phi)\La^{2}}\, ,
\end{equation}
We call this the ``B scheme''.
Then inserting in \eqref{eq:genPT_flow} we arrive at the RG equation
%
%

%
%
\begin{equation} \label{eq:new-ERGEadj}
 \La \frac{d}{d \La} S_{\La}[\varphi]
 = \Tr \left[ \left( 1 + \frac{1}{2} \La \frac{d}{d \La} \log Z_{\La} \right)
 \left( \frac{ m \La^{2} Z_\La(\phi)}{ S^{(2)}_{\La}[\varphi] + m \La^{2}Z_\La(\phi)} \right)^{m} \right].
\end{equation}
In the limit $m \to \infty$ one gets
\begin{equation} \label{eq:expadj}
 \La \frac{d}{d \La} S_{\La}[\varphi]
 = \Tr \left[ 
\left(1+\frac{1}{2} \La \frac{d}{d \La} \log Z_\La(\phi) \right)
\ e^{- \frac{S^{(2)}_{\La}[\varphi]}{\La^{2}Z_\La(\phi)}} \right].
\end{equation}
%

In the preceding discussion we have assumed that the
``spectral adjustment'' is introduced at the level of the
cutoff function $\rho_{k,\La}$.
In principle, it could also
be carried out at the level of the infinitesimal RG transformation
\eqref{eq:cut}, or (\ref{eq:spectral_der_rho_gamma}), namely
\begin{equation} \label{eq:spectral_der_rho_gammaC}
r_{\La}(s ; m)
= \frac{2}{\G(m)}(m s Z_\La(\phi)\La^{2})^{m} e^{- m s \La^{2}}\, .
\end{equation}
This then leads to an RG equation that is similar to \eqref{eq:expadj}, but without the term $\La \frac{d}{d \La} \log Z_\La(\phi)$.
We will refer to it as the ``C scheme''.

%
\section{Relation to formal coarse-graining schemes}
%
The main goal of this Section is to investigate in a formal way if  the family of flow equations
Eq.~(\ref{eq:genPT_flow}), 
can be interpreted as Wilsonian flows.
In general terms the coarse-graining procedure for the Wilsonian action
can be defined as
\begin{equation} \label{eq:generalCG}
 \La \frac{d}{d \La} e^{-S_\La[\varphi]}
 = \int \! d x \, \frac{\delta}{\delta  \varphi(x)} \left( \psi^{\La}_{x}[\varphi] e^{- S_{\La}[\varphi]} \right)\ ,
\end{equation}
for some $\psi^\La_{x}[\varphi]$.
This equation implies that the partition function 
$Z = \int [d \varphi] e^{-S_{\La}[\varphi]}$ 
is manifestly independent of $\Lambda$.
The specific form of the quantity in brackets leads to
the following general form for Wilsonian flow:
\begin{equation} 
\label{eq:generalFlow}
 \La \frac{d}{d \La} S_{\La}[\varphi]
 = \int d x \left( \frac{\delta S_{\La}[\varphi]}{\delta \varphi(x)} \psi^{\La}_{x}[\varphi]
 - \frac{\delta \psi^{\La}_{x}[\varphi] }{ \delta \varphi(x)} \right) \, .
\end{equation}
We also recall that this general Wilsonian flow is associated to an infinitesimal field redefinition
$\varphi(x) \to \varphi'(x) = \varphi(x) - \frac{\delta \La}{\La} \psi_{x}^{\La}[\varphi]$.
%
%
Indeed starting from a given coarse-graining map $\varphi \to b_\La(\varphi)$, transforming the field $\varphi_0$ at the scale $\La_0$ into the field $\varphi$ at the scale $\La<\La_0$, one can derive
the following relation \cite{Latorre:2000qc,Rosten}
\begin{equation}
 \psi_{x}^{\La}[\varphi]
 = e^{S_{\La}[\varphi]} \int [d \varphi_{0}] \, \delta( \varphi - b_{\La}[\varphi_{0}] ) \,
 \La \frac{ d b_{\La}[\varphi_{0}](x) }{ d \La } \, e^{- S_{\La_{0}}[\varphi_0]} \, .
\end{equation}
It is then natural to ask whether the general flow
equation \eqref{eq:generalFlow} can be reduced to the PT flow in Eq.~(\ref{eq:genPT_flow}),
and if so, what conditions should $\psi^\La_{x}[\varphi]$ satisfy.

Let us first recall what happens in the case of the Wilsonian flow 
considered by Polchinski \cite{Polchinski_1984}.
In this case one can directly guess the form of $\psi_{x}^{\La}[\varphi]$ to be plugged in the
general formula~(\ref{eq:generalFlow}):
\begin{equation}
 \psi_{x}^{\La}[\varphi]
 = \frac{1}{2} \int \! d z \, \dot{\Delta}_{x z} \, \frac{ \delta \Sigma_\La[\varphi] }{ \delta \varphi(z)} \, ,
 \label{polc_red}
\end{equation}
where $\Delta$ is a suitably regulated propagator with a dot standing for the derivative w.r.t.~$\log \La$ and
$\Sigma_\La[\varphi]$ is given by
\begin{equation}
 \Sigma_{\La}[\varphi] = - \frac{1}{2} \int \! d x \, \varphi(x) (-\Box_{x}) \varphi(x) + S^{I}_{\La}[\varphi] \, , \quad
 S_{\La}[\varphi] = \frac{1}{2} \int \! d x \, \varphi(x) (-\Box_{x}) \varphi(x) + S^{I}_{\La}[\varphi] \, .
 \label{polc_def}
\end{equation}
This choice leads to the desired flow equation
\begin{equation} \label{eq:polchinski_eq}
 \La \frac{d}{d \La} S^{I}_{\La}[\varphi]
 = \frac{1}{2} \int \! d x d y \, \dot{\Delta}_{x y}
 \left[ \frac{\delta S_\La^{I}[\varphi]}{\delta  \varphi(y)} \frac{\delta S_\La^{I}[\varphi]}{\delta  \varphi(x)}
 - \frac{\delta^{2} S_\La^{I}[\varphi]}{\delta  \varphi(y) \delta \varphi(x)} \right] \, .
\end{equation}
The same RG flow equation can be actually obtained also by other choices of $\psi_{x}^{\La}[\varphi]$, belonging to an infinite family obtained adding to the expression in Eq.~\eqref{polc_red}
a term $w_x[\varphi] e^{s_\La[\varphi]}$, such that $ \int \! d x \, \frac{\delta}{\delta  \varphi(x)} w_{x}[\varphi] = 0$. They correspond to different (but equivalent, in generating the flow) implementations of the coarse-graining procedure.

The Polchiski Wilsonian effective action $S_\La$, which satisfies the above flow equation, gets, in general, both 1PI (the second term) and 1PR (the first term) contributions.
It is known in this case that only in the sharp cutoff limit, if the momenta flowing into a vertex of a 1PR term have sum below the UV cutoff $\La$, then such contributions are absent~\cite{Morris:1993qb}.
Similar considerations are valid for the Wegner-Houghton Wilsonian action.
Moreover the Polchinski action has a simple relation with the regulated generator of the connected Green's functions and with the effective average action~\cite{Morris:1993qb}.

Going back to the goal of this section, in order to interpret in the Wilsonian sense the general PT flow Eq.~(\ref{eq:genPT_flow}),
we would have to rewrite it in the form given in the Eqs.~(\ref{eq:generalFlow})
or~(\ref{eq:generalCG}).
Thus we must look for a solution of the following functional equation
\begin{equation} \label{eq:condpsi2}
 \int \! d x \, \frac{\delta}{\delta \varphi(x)} \left( \psi^\La_{x}[\varphi] e^{- S_{\La}[\varphi]} \right)
 = - e^{- S_{\La}[\varphi]} \frac{1}{2} \tr \int_{0}^{\infty} \frac{d s}{s}
 \left[ r_{\La}(s) e^{-s S^{(2)}_{\La}[\varphi]} \right] \, .
\end{equation}
Before moving to this task let us make a comment. The existence of a solution would make it possible to interpret this particular PT regulated action $S_\La$, as a Wilsonian action, i.e. an action which, inserted in a functional integral,
not only generates the partition function $Z$, but also all the possible correlators (connected and not connected) with momenta below the scale $\La$, i.e.
\begin{equation}
\langle O_1(x_1) \cdots O_n(x_n) \rangle=\frac{1}{Z} \int [d \varphi]_\Lambda e^{-S_\La[\varphi]} O_1(x_1) \cdots O_n(x_n) \,.
\end{equation}
From the structure of the flow equation one notes that in general, contrary to the Polchinski action, this action gets along the flow contributions which results into 1PI non local vertices. 
We stress that the relation between the PT regulated $S_\La$ and the effective action $\Gamma$ (eventually IR regulated) 
is not trivial and certainly not so simple as for the Polchinski Wilsonian action.
%
\subsection{Construction of a general solution}
%
Let us then ask the following question:
given a certain Wilsonian RG flow is it possible to reconstruct the associated
coarse-graining procedure which may be encoded in the infinitesimal generator $\psi^\La_x[\varphi]$?
We have already seen for the specific case of a Polchinski RG flow that this procedure is not unique.
From the definition recalled above it is clear that such a construction requires knowledge of the 
Wilsonian action $S_{\La}[\varphi]$
satisfying the RG flow equation, making this problem hard to solve in practice.
Nevertheless it is important to know if at least the solution of this problem can exist.

The left hand side of this equation can be seen as a divergence of a vector field belonging to an infinite dimensional vector
space, i.e., has the structure
\begin{equation} \label{infdiv}
 \int \! d x \, \frac{\delta}{\delta \varphi(x)} u_{x}[\varphi] = f[\varphi] \, .
\end{equation}
Keeping in mind that the quadratic kinetic part of the Wilsonian action is defined by a UV regulated Laplacian
$(\Box)_\Lambda$, it is convenient to define a slightly different object, which already simplifies the analysis for the free theory case as discussed in the next subsection,%
\footnote{%
We thank Tim Morris for this suggestion.
}
\begin{equation}
 u_{x}[\varphi] = G_{x y} v_{y}[\varphi]
\end{equation}
where we have introduced the space-time 'regulated' Green's function $G_{x y}$ satisfying
\begin{equation}
 (-\Box_{x})_{\La} G_{x y} = \delta(y) \, .
\end{equation}

One can assume that as in any finite dimensional vector space, the vector field $v_{x}[\varphi]$ can be decomposed as a sum
of a ``gradient'' part and a divergenceless part as follows
\begin{equation} \label{decomp2}
 v_{x}[\varphi]
 = \frac{\delta}{\delta \varphi(x)} h[\varphi] + (-\Box_{x})_{\La} w_{x}[\varphi], \quad
 \int \! d x \, \frac{\delta}{\delta  \varphi(x)} w_{x}[\varphi] = 0 \, .
\end{equation}

Then a particular solution of Eq.~(\ref{infdiv}) using Eq.~(\ref{decomp2}) can be obtained solving an infinite dimensional
Poisson-like equation
\begin{equation} \label{infpoisson2}
 \int \! d x d y \, \frac{\delta}{\delta \varphi(x)} G_{x y} \frac{\delta}{\delta \varphi(y)} h[\varphi] = f[\varphi] \, ,
\end{equation}
where the metric in field space is field independent (flat geometry in field space) even if has a non trivial space-time
dependence.
Since the determinant of the metric is field independent one can see that the operator in field space above is really a
covariant Laplacian.
We expect that this generalized ``elliptic'' second order linear differential problem can have solutions,
unique at least from the physical point of view when suitable boundary conditions are imposed. 

Let us write a functional Fourier transform for the scalar $h$ introducing a source $J$
\begin{equation}
 \tilde h[J] = \int [d \varphi] e^{-i \varphi \cdot J} h[\varphi]
\end{equation}
and similarly for the right hand side of Eq.~(\ref{infpoisson2}), i.e. for $f[\varphi]$.
Then one can formally rewrite this generalized Poisson equation as
\begin{equation} \label{Fourierpoisson1}
 - (J \cdot G \cdot J) \, \tilde h[J] = \tilde f[J] \, ,
\end{equation}
where $J \cdot G \cdot J = \int \! d x d y \, J (x) G_{x y} J(y)$, and derive its solution
\begin{equation}
 h[\varphi] = - \int [d J] e^{i \varphi \cdot J} \frac{1}{J \cdot G \cdot J} \tilde f[J] \, .
\end{equation}

Using this setup for our original problem we can then formally write
\begin{equation} \label{approach2}
 \psi^\La_{x}[\varphi]
 = e^{S_{\La}[\varphi]} \left( \int [d J] e^{i \varphi \cdot J} \frac{- i J(x) }{J \cdot G \cdot J} \tilde f[J]
 + w_{x}[\varphi] \right)\,,
\end{equation}
where $w_{x}[\varphi]$ is an arbitrary divergenceless vector field according to Eq.~(\ref{decomp2}),
which again can be chosen to improve the behavior of the solution.
Therefore we can see that formally, given a Wilsonian flow, one can construct a coarse-graining procedure which generates it.
Moreover the coarse-graining procedure is not unique, indeed there are infinitely many. One can take eventually advantage
of this freedom to make the most sensible choice of coarse-graining from the physical point of view.

One can even try to formalize this picture considering the space of Wilsonian actions ${\cal S}$ and the space of
Wilsonian RG flows ${\cal V}$, where each point $p \in {\cal V}$ is the vector field functional of the flow, e.g.,
associated to Eq.~(\ref{eq:generalFlow}).
Then one can consider a fiber bundle space ${\cal F}$ such that, to each point $p$ belonging to its base ${\cal V}$ is
associated a fiber related to the coarse-graining generator $\psi^\La_{x}$.
Like in a gauge theory an infinite set of $\psi^\La_{x}$ is associated to the same vector field $p$
generating the Wilsonian RG flow.
%

In summary at least formal solutions of Eq.~(\ref{eq:condpsi2}) can be constructed.
This is true also in the spectrally adjusted cases named B and C in the end of Section~\ref{sec2}, 
which are the ones which make it possible to interpret a proper time coarse-graining in terms of a momentum coarse-graining in the derivative expansion.  
This seems to confirm that the general PT regulated flow equation~(\ref{eq:genPT_flow}) can represent a Wilsonian RG flow.
%
%
\subsection{The free theory case}
%
Given that a formal solution can be constructed, one may ask which kind of requirements should be further imposed
on the field redefinitions by the Wilsonian coarse graining procedure from a reasonable physical point of view. 
Clearly one would like to avoid possible pathological definitions.
This question can be posed in general and in particular for the specific case of a proper time flow.
First one can notice that even in the Polchinski flow for a free theory, according to Eqs.~\eqref{polc_red} and~\eqref{polc_def}, there is some degree of non locality in the redefinition of the fields,
which depends on the regulator properties encoded in $\Delta$.
Further non local behavior, at least in the IR regime, is expected in presence of new degrees of freedom in the spectrum of the theory, such as bound states.

We shall then investigate what happens for the case of a proper-time flow, but which could hold also for more general class of flows.
Let us then investigate in detail the case of a free quadratic action for the case of a generic cutoff
\begin{equation}
 S[\varphi] = \frac{1}{2} \varphi \cdot (-\Box)_{\Lambda} \cdot \varphi , \quad
 f[\varphi] = - \frac{1}{2} e^{-S_{\La}[\varphi]} \, {\rm Tr} \int_{0}^{\infty} \frac{d s}{s}
 \left[ r_{\La}(s) e^{-s (-\Box)_{\La}} \right]
\end{equation}
so that one can write
\begin{equation}
 \tilde f[J]
 = - \frac{1}{2} {\rm Tr} \int_{0}^{\infty} \frac{d s}{s}
 \left[ r_{\La}(s) e^{- s (- \Box)_{\La}} \right] \left( {\rm Det} G\right)^{\frac{1}{2}} e^{- \frac{1}{2} J \cdot G \cdot J} \, .
\end{equation}
Using Eq.~(\ref{approach2}) we can then write the solution for the ``potential'' $h$ as
\begin{equation} \label{sol2}
 h[\varphi]
 = \frac{1}{2} {\rm Tr} \! \int_{0}^{\infty} \frac{d s}{s} \!
 \left[ r_{\La}(s) e^{- s (- \Box)_{\La}} \right]
 \int [d J] e^{i \varphi \cdot J} \frac{\left( {\rm Det} G\right)^{\frac{1}{2}}}{J \cdot G \cdot J} e^{- \frac{1}{2} J \cdot G \cdot J}
 = {\cal N} \!\! \int [d \tilde{J}] e^{i  \tilde{J} \cdot G^{-\frac{1}{2}} \varphi}
 \frac{1}{ \tilde{J} \cdot \tilde J} e^{- \frac{1}{2} \tilde{J} \cdot \tilde{J}} ,
\end{equation}
where $\tilde{J} = G^{\frac{1}{2}} J$ and we have defined the normalization factor
\begin{equation}
 {\cal N}
 = \frac{1}{2} {\rm Tr} \int_{0}^{\infty} \frac{d s}{s} \left[ r_{\La}(s) e^{- s (- \Box)_{\La}} \right]\,.
\end{equation}
One has to keep in mind that a functional derivative must be taken to construct $v_{x}$ and the infinitesimal Wilsonian
field redefinition $\Psi^\La_{x}$ as well, so that in the integral in Eq.~(\ref{sol2}) the singular region in the origin (in $\tilde{J}$ space) is harmless and
one could replace $e^{i \tilde{J} \cdot G^{- \frac{1}{2}} \varphi} \to \left( e^{i \tilde{J} \cdot G^{-\frac{1}{2}} \varphi }
- 1 - i \tilde{J} \cdot G^{- \frac{1}{2}} \varphi \right)$, given that non zero contributions in the integral in $\tilde{J}$
comes from even integrands. One can then perform the Fourier transform.

We follow here an alternative path in the computation by introducing a parameter $a$ to be set to $\frac{1}{2}$ at the end in the previous formal solution for $h[\varphi]$:
\begin{equation} \label{sol2a}
 I_{a}[\varphi]
 = {\cal N} \int [d \tilde{J}] e^{i \tilde{J} \cdot G^{-\frac{1}{2}} \varphi}
 \frac{1}{\tilde{J} \cdot \tilde{J}} e^{- a \tilde{J} \cdot \tilde{J}} ,
\end{equation}
and, after taking a derivative w.r.t.~$a$, perform the $\tilde{J}$ functional integration
\begin{equation} \label{sol2bis}
 - \frac{d}{d a} I_{a}[\varphi]
 = {\cal N} \int [d \tilde{J}] e^{i \tilde{J} \cdot G^{- \frac{1}{2}} \varphi} e^{- a \tilde{J} \cdot \tilde{J}}
 = {\cal N} \frac{\sqrt{\pi}}{\sqrt{a}} e^{- \frac{ \varphi\cdot (- \Box)_{\La}\cdot \varphi}{4 a}} .
\end{equation}
Therefore $I_a$ can be thought of as a function of $S_\Lambda[\varphi]=\frac{1}{2}\varphi \cdot (-\Box)_{\La} \cdot \varphi$.
Re-integrating back in $a$ one finds
\begin{equation}
 I_{a}[\varphi]
 = - {\cal N} \left[ 2 e^{- \frac{\varphi \cdot (- \Box)_{\La}\cdot  \varphi}{4 a} } \sqrt{\pi a}
 + \pi \sqrt{\varphi \cdot (- \Box)_{\La}\cdot  \varphi} \,\, {\rm erf} \left( \sqrt{ \frac{\varphi \cdot(-\Box)_{\La} \cdot\varphi}{4 a} } \right)
 + c[\varphi] 
 \right] \, ,
\end{equation}
where the last term is an $a$-independent functional. From a direct integration it turns out to be a simple constant
and can be discarded.
Setting $a = 1/2$ and ignoring also $w_{x}[\varphi]$ in Eq.~(\ref{decomp2}) one gets
\begin{eqnarray}
 \nonumber
 \!\!\!\!\!\! \psi_{x}[\varphi]
 \!\!\!&=& e^{-\frac{1}{2} \varphi \cdot(-\Box)_{\La}\cdot \varphi}  G_{x y} \frac{\delta h[\varphi]}{\delta \varphi(y)}
 = - {\cal N} e^{-\frac{1}{2} \varphi \cdot(-\Box)_{\La} \cdot \varphi} \frac{\pi}{ \sqrt{ \varphi \cdot(-\Box)_{\La}\cdot \varphi } }
 {\rm erf} \left( \sqrt{ \frac{\varphi \cdot(- \Box)_{\La}\cdot \varphi}{2} } \right) \varphi_{x}
 \\
 &=& \!\!- {\cal N} \sqrt{2 \pi} \left( 1 - \frac{2}{3} \varphi \cdot(- \Box)_{\La}\cdot \varphi
 + \frac{7}{30} ( \varphi\cdot (- \Box)_{\La}\cdot \varphi )^{2} - \frac{2}{35} ( \varphi\cdot (- \Box)_{\La} \cdot\varphi)^{3} + \cdots
 \right) \varphi_{x} \, .
\end{eqnarray}

\noindent
We note that in this free theory the coarse-graning is, as said, a function of the free Wilsonian action, and therefore non local, however this non locality is harmless in this case.

More generally, in presence of interactions, any analysis is extremely complicated and it could be carried on only with drastic simplifications.
A fortiori in this case we should not expect in general quasi locality in the action, e.g. in strongly interacting theories where bound states appear in the spectrum.
In such cases the flow equation is also bound to be non-local.
We note, however, that infinitely many coarse-graining schemes give the same flow and that
the freedom to introduce the divergenceless vector in field space,
as defined in Eq.~(\ref{decomp2}) may be helpful to eliminate some pathological
non-localities.
Similar considerations can be applied in a quasi local regime, i.e. when the action can be written as a power series in derivatives.

%
%
\section{Working example: the Wilson-Fisher fixed point}\label{ising}

It is interesting to  apply this formalism to the determination of critical exponents for the Wilson-Fisher fixed point.
Let us therefore consider the following euclidean action,
which represents the leading term in a derivative expansion:
\be\label{action}
S_\Lambda[\phi]= \int d^d x \left ( \frac{1}{2}\ZL \partial_\mu \phi \partial^\mu \phi + \VL \right ) ,
\ee
We will study the flow of this action with a modified type-A regulator.
In order to take into account the anomalous dimensions,
we shall assume that the field does not have the standard mass
dimension $d/2-1$ but rather $d/2-1+\eta/2$.
Then the prefactor $\ZL$ has dimension $-\eta$.
Accordingly, we modify the cutoffs of equations
\eqref{eq:spectral_rho_gamma} and
\eqref{eq:spectral_der_rho_gamma}
to
\begin{align}
\rho_{k, \La}^A(s ; m) ={}& \frac{\G(m,msk^{2-\eta}) 
- \G(m , ms\La^{2-\eta})}{\G(m)} ,
\label{eq:param_spectral_rho_gamma}
\end{align}
and
\begin{equation}
\label{nonspec}
r^{A}_\Lambda = (2 - \eta) \frac{(m \, s \, \Lambda^{2-\eta})^m}{\Gamma(m)} e^{- m \, s \, \Lambda^{2 - \eta}}\ .
\end{equation}
We will call this a type-A' regulator.

Expressing the functional trace in terms of a momentum integral, the PT flow equation reads
\begin{equation}\label{pt1}
\Lambda\frac{d}{d\Lambda} S_\Lambda[\phi] = \frac{1}{2}
\int d^d x \int \frac{d^d p}{(2\pi)^d} \langle x | r_\Lambda \, e^{-s S_\La^{(2)}[\phi] } | p \rangle \langle p | x \rangle .
\end{equation}
This can be evaluated by employing the derivative expansion within the functional trace and then performing the momentum integral.
We will take into account all terms up to two derivatives, i.e., we will calculate the flow of $Z_{\La}(\phi)$ and $V_{\La}(\phi)$.
Thus, for evaluating the functional trace, we need to commute all derivatives to the right,
\bea \label{commutation}
[ \partial_{\mu} , f(\phi) ] = f'(\phi) (\partial_\mu \phi)
\eea 
in order to make use of the identity $\partial_{\mu} | p \rangle = i p_{\mu} | p_{\mu} \rangle$ and finally project the flow on the running of $Z_\Lambda(\phi)$  and $V_\Lambda(\phi)$. The details of the derivation as well as the equations are provided in the appendix, where the flow equations for the dimensionless versions of 
$Z_\La$ and $V_\La$ (denoted $z$ and $v$) are given in Eqs.~\eqref{fpv} and~\eqref{fpz}.

\begin{figure}[htpb]
\includegraphics[width=.6\textwidth]{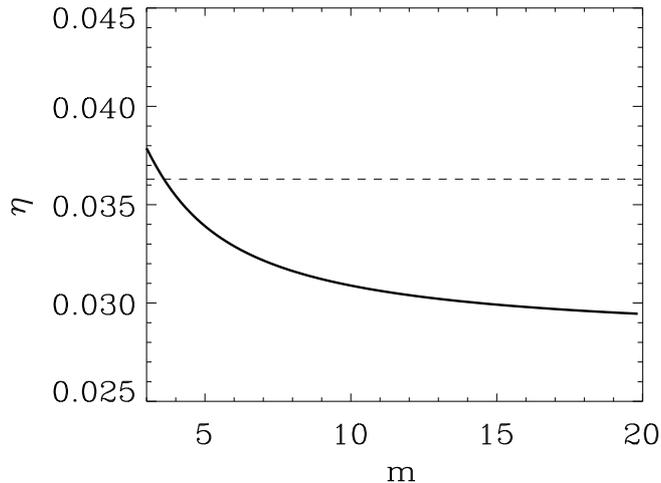}
\caption{The anomalous dimension $\eta$ for the  regulator considered (type-A) in this work as a function of the proper-time parameter $m$. 
In the $m=\infty$  limit we find $\eta=0.0294$. 
The dashed line indicates $\eta=0.03629$ obtained with conformal bootstrap method which is the most accurate determination nowadays. \label{critical}}
\end{figure}

The fixed-point solutions are obtained by means of  a multiple shooting method. In particular (see \cite{morris2} for an extended discussion), 
a large field expansion is  assumed to be valid at some fixed value of the field where the initial conditions are set, and  an inward numerical integration of
the fixed point equations (\ref{fpv}) and (\ref{fpz}) is performed.  At $x=0$ instead the initial condition for the outward integration is performed
assuming $z(x=0)=1$, $z'(x=0)=v'(x=0)=0$. For actual calculations the  Runge-Kutta integrator DO2PVF~%
\footnote{%
implemented by the \NAG group (see \url{https://www.nag.com} for details).
}
turned out to be rather efficient.  Continuity at the fitting point of the functions $Z$, $V$  and their derivatives  is obtained by means of a globally  convergent Broyden's method.%
\footnote{%
This is described for example in \url{http://numerical.recipes}.
}
In particular a tolerance of  $10^{-6}$ for the root finding algorithm  has been assumed. 
The result for the anomalous dimensions are depicted in Fig.~\ref{critical}.

The critical exponent $\nu$ is instead almost constant as a function of $m$ and its value is $\nu=0.613$.

Let us also mention the equations for the other types of coarse-graining schemes (type B and C) previously discussed. 
We shall derive the flow equations but postpone to another work the full numerical analysis for these schemes.
We introduce  a regulator, field dependent in general, of the form, cf.~Eq.~(\ref{eq:spectral_rho_gamma}),
\begin{align}
 r_{\La}^{p}(s;m)
 ={}& \big( 2 - \eta (1 - p) + \alpha \, p \, \La \partial_{\La} \ln Z_{\La}(\phi)^{p} \big) \frac{(m \, s \, Z_{\La}(\phi)^{p} \, \Lambda^{2 - \eta (1 - p)})^{m}}{\G(m)} e^{- m \, s \, Z_{\La}(\phi)^{p} \, \Lambda^{2 - \eta (1 - p)}} .
 \label{eq:param_spectral_der_rho_gamma}
\end{align}
%
The various classes of regulators can be distinguished in terms of the parameters $p$ and $\alpha$.
For $p = 0$ we have a regulator of type-$A$ while for $p = 1$ we have regulators of type B or C for $\alpha =1$ or $\alpha =0$, respectively.

In addition, in the type-$C$ regulator one can replace
\begin{align}
 \int \!\! {\rm d}^{d} x \, r_{\La}^{p}(\phi) \, f(\phi) \, ( \square \phi )
 \to  -\int \!\! {\rm d}^{d}x \, \big[ r_{\La}^{p}(\phi) \, f'(\phi) + 
\beta \, r_{\La}^{p^\prime} (\phi) \, f(\phi) \big] (\partial^{\mu} \phi) (\partial_{\mu} \phi) .
\end{align}
In the full type-$C$ case one should keep $\beta=1$. 
Historically, the PT flow equations used in the literature have been derived heuristically by a simple RG-improvement
of the 1-loop proper-time expression, 
\cite{Bonanno:2000yp,Mazza:2001bp,Litim:2010tt}
and that  simplified type-C scheme corresponds to setting $\beta=0$.
On the contrary, 
regulators of type $A$ and $B$ have never been used to calculate 
the critical exponents for the Wilson-Fisher fixed point.

The flow equations for $v$ and $z$ in these spectrally adjusted coarse-graining schemes are given in the Appendix in 
Eqs.~\eqref{flowv_gen} and~\eqref{flowz_gen} as a function of $p$, $\alpha$ and $\beta$.
%
\section{Conclusions}

We have argued that, at least at a formal level, 
the PTRG equation \eqref{eq:PTexp}
and its generalization \eqref{eq:genPT_flow} can indeed be seen
as Wilsonian RG equations.
In fact, we found that each such flow can be associated to an infinite family of coarse-graining procedures.
In practical applications, this freedom could be exploited to
avoid certain pathologies.

The definition of the flow depends, as usual, on several choices.
Foremost among these is the choice of a regulator function. Among all possible definitions,
we have considered certain regulators based on
incomplete Gamma functions.
Then, we have considered the freedom of introducing 
a ``spectral adjustment'' either in the cutoff itself or in its
derivative with respect to $\log\Lambda$. 
These spectral choices have the  advantage of making possible a more direct interpretation for the coarse-graining as a suppression of the modes with $p^{2} > \La^{2}$.
We have implemented the flow equation
of the Ising universality class, using a regulator without spectral adjustment (type-A).
Depending on the value of the parameter $m$ the results for the anomalous dimension
$\eta$ and for the exponent $\nu$ in the scheme-A are actually of a quality a bit lower compared to other RG equations.
From a preliminary analysis they are improving moving to the spectrally adjusted RG equations (type B and C). Such results
will be discussed elsewhere.
\bigskip

{\it Acknowledgments:}
We thank Tim Morris, Jan Pawlowski and Kevin Falls 
for interesting discussions.
S.L.~is supported by the DFG Collaborative Research Center "SFB 1225 (ISO-QUANT)".

\section*{Appendix: Derivative expansion}
In this appendix the main steps to obtain the flow equation in the derivative expansion 
are given. As this involves quite some algebra, we use the Mathematica package xAct~\cite{xa2,xactw}.
However, I 
In a first step we split the second functional derivative into two pieces,
\begin{align} \label{split}
S^{(2)}_{\La}[\phi] ={}& S^{(2)}_{0}[\phi] + \DS[\phi],
\end{align}
the first contains the box operator,
\begin{align}
S^{(2)}_0 [\phi] ={}& \ZL \square ,
\end{align}
and the second is the remainder,
\begin{align}
\DS[\phi] = -Z_\Lambda'(\phi) (\partial^\mu \phi) \, \partial_\mu +
Z_\Lambda'(\phi) (\square \phi) - \frac{1}{2} Z_\Lambda''(\phi) (\partial_\mu \phi) (\partial^\mu \phi)
+ V_\Lambda''(\phi).
\end{align}
This split allows us to only take into account finitely many commutators when evaluating the exponential in Eq.~\eqref{pt1},
as higher orders lie outside of our truncation.
Using the Zassenhaus (Baker-Campbell-Haussdorff) formula we find
\begin{align} \label{eq:Zassenhaus}
e^{-s S_\La^{(2)}[\phi] }
= e^{\frac{s^{3}}{3} [ \DS[\phi] , S_{0}^{(2)}[\phi] ]_{2} - \frac{s^{3}}{6} [S_{0}^{(2)}[\phi] , \DS[\phi] ]_{2} } e^{\frac{s^{2}}{2} [S_{0}^{(2)}[\phi] , \DS[\phi] ]}
e^{- s \DS[\phi]}
e^{-s S_{0}^{(2)}[\phi]}
+ \mathcal{O}(\partial^{3} \phi),
\end{align}
where $[X,Y]_{n+1} = \big[ X, [X,Y]_{n} \big]$ and $[X,Y]_{0} = Y$.
The first two exponentials can be expanded to finite order, as they contain at least one derivative of the field.
For the third exponential, we perform a second split.
We rewrite $\DS[\phi]$ as,
\begin{align}
\DS[\phi] = \DS_{\partial}[\phi] + \DS_{0}[\phi],
\end{align}
where $\DS_{\partial}[\phi]$ contains an actual operator,
\begin{align}
\DS_{\partial}[\phi] = -Z_\Lambda'(\phi) (\partial^\mu \phi) \, \partial_\mu,
\end{align}
while $\DS_{0}[\phi]$ is just a number,
\begin{align}
\DS_{0}[\phi] =
Z_\Lambda'(\phi) (\square \phi) - \frac{1}{2} Z_\Lambda''(\phi) (\partial_\mu \phi) (\partial^\mu \phi)
+ V_\Lambda''(\phi) .
\end{align}
So we can use the Zassenhaus Formula again,
\begin{align}
e^{- s \DS[\phi]}
= e^{- \frac{s^{2}}{2} [\DS_{\partial}[\phi], \DS_{0}[\phi]]} e^{- s \DS_{\partial}[\phi]} e^{- s \DS_{0}[\phi]} + \mathcal{O}(\partial^{3} \phi),
\end{align}
and as before, we can expand the first two exponentials to finite order.
Finally, we need to evaluate the most right exponent in \eqref{eq:Zassenhaus}, $e^{- s S_{0}^{(2)}}$.
For this let us decompose $\big(Z_\La(\phi) \square \big)^{n}$,
\begin{align}
\big( Z_\La(\phi) \square \big)^{n}
= \sum_{i=0}^{4} a_{n}^{i} \, I_{i} \, Z_\La(\phi)^{n-m_{i}} \square^{n-m_{i}} + \mathcal{O}(\partial^{3} \phi),
\end{align}
in terms of the invariants $I_{i}$ containing up to two derivatives of the field,
\begin{align}
\notag
{}& I_{0} = \mathbbm{1}, &
\quad
{}& \!\!\!\!\!\!\! I_{1} = [\square, \! Z_\La(\phi)], &
\quad
{}& \!\!\!\!\!\!\! I_{2} = [Z_\La(\phi), \! \square]_{2} \square, &
\quad
{}& \!\!\!\!\!\!\! I_{3} = [\square, \! Z_\La(\phi)]^{2} Z_\La(\phi) \square, &
\quad
{}& \!\!\!\!\!\!\! I_{4} = [\square, \! Z_\La(\phi)]_{2} Z_\La(\phi),
\\
{}& m_{0} = 0, &
\quad
{}& \!\!\!\!\!\!\! m_{1} = 1, &
\quad
{}& \!\!\!\!\!\!\! m_{2} = 2, &
\quad
{}& \!\!\!\!\!\!\! m_{3} = 3, &
\quad
{}& \!\!\!\!\!\!\! m_{4} = 2.
\end{align}
By induction one can show that the $a_{n}^{i}$ satisfy the recursion relation,
\begin{align}
\notag
{}& a_{n+1}^{0} = 1, &
\qquad
{}& a_{n+1}^{1} = a_{n}^{1} + n, &
\qquad
{}& a_{n+1}^{2} = - a_{n}^{1} + a_{n}^{2} - \frac{n(n+1)}{2}, &
\\
{}& a_{n+1}^{3} = (n-1) a_{n}^{1} + a_{n}^{3}, &
\qquad
{}& a_{n+1}^{4} = a_{n}^{1} + a_{n}^{4}, &
\end{align}
with the initial conditions
\begin{align}
a_{0}^{0} = a_{1}^{0} = 1,
\qquad
a_{0}^{i > 0} = a_{1}^{i > 0} = 0.
\end{align}
The solution of this recursion reads
\begin{align}
\notag
{}& a_{n}^{0} = 1, &
\quad
{}& a_{n}^{1} = \frac{n(n-1)}{2}, &
\quad
{}& a_{n+1}^{2} = -\frac{n(n-1)(2n-1)}{6}, 
\\
{}& a_{n+1}^{3} = \frac{n(n-1)(n-2)(3n-5)}{24}, &
\quad
{}& a_{n+1}^{4} = \frac{n(n-1)(n-2)}{6}. &
\end{align}
After determining the $a_{n}^{i}$ we can evaluate the exponential,
\begin{align}
\notag
e^{- s S_{0}^{(2)}}
={}& \sum_{n=0}^{\infty} \frac{(-s)^{n}}{n!} \big(Z_\La(\phi) \square \big)^{n}
= \sum_{n=0}^{\infty} \frac{(-s)^{n}}{n!}
 \sum_{i=0}^{4} a_{n}^{i} \, I_{i} \, Z_\La(\phi)^{n-m_{i}} \square^{n-m_{i}} + \mathcal{O}(\partial^{3} \phi)
 \\
 ={}& \sum_{i=0}^{4} I_{i} \sum_{n=0}^{\infty} \frac{(-s)^{n}}{n!}
 a_{n}^{i} \, Z_\La(\phi)^{n-m_{i}} \square^{n-m_{i}} + \mathcal{O}(\partial^{3} \phi),
\end{align}
by using the identity
\begin{align}
\sum_{n=0}^{\infty} \frac{n^{m}}{n!} x^{n} = (x \partial_{x})^{m} e^{x} .
\end{align}

At last, the dimensionless quantities  $x= \phi \Lambda^{-\frac{d-2+\eta}{2}}$ ,  
$v=V \Lambda^{-d}$ and $z=Z \Lambda^{\eta}$ where $\eta$ is the anomalous dimension, are introduced. 
For a type-$A$ cutoff ($p=0$) the flow equations for $v$ and $z$ read
\begin{eqnarray}
\label{fpv}
&&\dot{v}=d v-\frac{1}{2} x (d+\eta -2) v'+\frac{2^{-d-1} \pi ^{-d/2} (\eta -2) m^m z^{-d/2} \Gamma\left(m-\frac{d}{2}\right) \left(m+v''\right)^{\frac{1}{2} (d-2m)}}{\Gamma (m)}
\end{eqnarray}
\begin{eqnarray}
\label{fpz}
&&\dot{z} =
-z \eta  -\frac{1}{2} x (d+\eta -2) z'-2^{-d-4} 
\pi ^{-d/2} (\eta -2) m^m z^{-\frac{d}{2}-1} \frac{\Gamma (1-\frac{d}{2}+m)}{3\Gamma(m)} \nonumber\\
&& \left (\left(v^{(3)}\right)^2 z^2 (d-2 (m+1)) (d-2 (m+2))+\left(m+v''\right)^2\left(((d-18) d-4) \left(z'\right)^2+24 z z''\right) \right. \nonumber\\
&& \left. -2 (d-10) v^{(3)} z (d-2 (m+1)) z' \left(m+v''\right) \right)
   \end{eqnarray}
   
For type-B and type C cutoff ($p=1$) the flow equations for $v$ and $z$ are significantly more involved. Unified formulae read
\begin{dmath}
\dot{v} = dv
-\frac{1}{2} x (d+\eta
   -2) v'+\frac{\gamma}{z}
 \left(\frac{z}{m
   z+v''}\right)^{m-\frac{d}{2}}
   \left(\alpha  x (d+\eta -2) z'+2 z
   (\alpha  \eta -2)+2 \alpha 
   \dot{z}\right)
   \label{flowv_gen}
\end{dmath}
where $\gamma= m^m \Gamma[m-\frac{d}{2}]/\Gamma[m]\pi^{d/2}2^{2+d}$. The flow equation for $z$ instead reads
\begin{dmath}
\dot{z}=\beta  2^{-d-4} \delta 
   z^{-\frac{d}{2}+m-2} \left(m
   z+v''\right)^{\frac{1}{2} (d-2 (m+3))}
   \left(z^2 \left(\alpha  \left(d^2-14
   d+40\right) m^2 x (d+\eta -2)
   \left(z'\right)^3+2 \alpha  m \dot{z}
   z' \left(\left(d^2-14 d+40\right) m
   z'-(d-6) v^{(3)} (d-2 (m+1))\right)-2
   \alpha  v^{(3)} (d-2 (m+1)) v''
   \left(x (d+\eta -2) z''+2
   \dot{z}'\right)+m \left(z'\right)^2
   \left(2 (d-10) v'' (-4 \alpha +d
   (\alpha  (\eta +2)-2)+2 m (\alpha 
   \eta -2)+4)-\alpha  (d-6) v^{(3)} x
   (d+\eta -2) (d-2 (m+1))\right)+2 v''
   z' \left(v^{(3)} (-d+2 m+2) (\alpha 
   (d+\eta -2)+2 m (\alpha  \eta -2))+2
   \alpha  (d-10) m x (d+\eta -2) z''+4
   \alpha  (d-10) m
   \dot{z}'\right)\right)+z v'' z'
   \left(2 \alpha  \dot{z} \left(2
   v^{(3)} \left(-d m+d+2
   m^2-2\right)+(d-10) m (d+2 m-6)
   z'\right)+2 z' \left(\alpha  (m-1)
   v^{(3)} x (d+\eta -2) (-d+2
   m+2)+(d-10) v'' (\alpha  (d+\eta -2)+2
   m (\alpha  \eta -2))\right)+\alpha 
   (d-10) m x (d+\eta -2) (d+2 m-6)
   \left(z'\right)^2+2 \alpha  (d-10) v''
   \left(x (d+\eta -2) z''+2
   \dot{z}'\right)\right)+2 m z^3
   \left(-\alpha  v^{(3)} (d-2 (m+1))
   \left(x (d+\eta -2) z''+2
   \dot{z}'\right)+z' \left(-v^{(3)} (d-2
   (m+1)) (-\alpha  (3 \eta +2)+d (\alpha
    \eta +\alpha -2)+8)+\alpha  (d-10) m
   x (d+\eta -2) z''+2 \alpha  (d-10) m
   \dot{z}'\right)+(d-10) m
   \left(z'\right)^2 (-\alpha  (\eta
   +2)+d (\alpha  \eta +\alpha
   -2)+4)\right)+2 \alpha  (d-10) (m-1)
   \left(v''\right)^2 \left(z'\right)^2
   \left(x (d+\eta -2) z'+2
   \dot{z}\right)\right)-2^{-d-5} \delta 
   z^{-\frac{d}{2}+m-2} \left(m
   z+v''\right)^{\frac{1}{2} (d-2 (m+3))}
   \left(z^2 \left(\left(v^{(3)}\right)^2
   \left(d^2-2 d (2 m+3)+4 \left(m^2+3
   m+2\right)\right)+\left(d^2-18
   d-4\right) m^2 \left(z'\right)^2-2
   (d-10) m v^{(3)} (d-2 (m+1)) z'+48 m
   v'' z''\right)+2 z v''
   \left(\left(d^2-18 d-4\right) m
   \left(z'\right)^2-(d-10) v^{(3)} (d-2
   (m+1)) z'+12 v''
   z''\right)+\left(d^2-18 d-4\right)
   \left(v''\right)^2
   \left(z'\right)^2+24 m^2 z^3
   z''\right) \left(\alpha  x (d+\eta -2)
   z'+2 z (\alpha  \eta -2)+2 \alpha 
   \dot{z}\right)-\frac{1}{2} x (d+\eta
   -2) z'-\eta  z
    \label{flowz_gen}
\end{dmath}
where $\gamma= m^m \Gamma[m-\frac{d}{2}]/\Gamma[m]\pi^{d/2}2^{2+d}$ and
$\delta=m^m \Gamma[1-d/2+m]/\Gamma[m] 3 \pi^{d/2}$. 
Setting $\alpha=\beta=1$, $\alpha=0$  and $\beta=1$, $\alpha=\beta=0$ one obtains schemes $B$, $C$ and simplified-C respectively, as discussed in the end of Section~\ref{ising}.



\end{document}